\documentclass[twocolumn,english,floats,showpacs,pre]{revtex4}
\usepackage[T1]{fontenc}
\usepackage[latin9]{inputenc}
\usepackage{textcomp}
\usepackage{amsmath}
\usepackage{amssymb}
\usepackage{graphicx}

\makeatletter

\newcommand{\lyxmathsym}[1]{\ifmmode\begingroup\def\b@ld{bold}
  \text{\ifx\math@version\b@ld\bfseries\fi#1}\endgroup\else#1\fi}

\@ifundefined{textcolor}{}
{%
 \definecolor{BLACK}{gray}{0}
 \definecolor{WHITE}{gray}{1}
 \definecolor{RED}{rgb}{1,0,0}
 \definecolor{GREEN}{rgb}{0,1,0}
 \definecolor{BLUE}{rgb}{0,0,1}
 \definecolor{CYAN}{cmyk}{1,0,0,0}
 \definecolor{MAGENTA}{cmyk}{0,1,0,0}
 \definecolor{YELLOW}{cmyk}{0,0,1,0}
}

\usepackage[above]{placeins}\usepackage{babel}

\usepackage{babel}
\usepackage{babel}
\usepackage{babel}
\usepackage{babel}
\usepackage{babel}

\usepackage{babel}

\usepackage{babel}

\makeatother

\usepackage{babel}
\begin{document}

\title{Short-range symmetry breaking induced structural collapse and $T_{c}$
enhancement in 122-type iron pnictide superconductors}

\author{X. Q. Huang$^{1,2}$}

\email{xqhuangnj@gmail.com}

\selectlanguage{english}%

\affiliation{$^{1}$Department of Telecommunications Engineering ICE, PLAUST,
Nanjing 210016, China \\
 $^{2}$ Department of Physics and National Laboratory of Solid State
Microstructure, Nanjing University, Nanjing 210093, China}

\date{\today}
\begin{abstract}
In this paper, we investigate analytically the experimental observations
of structural collapse and $T_{c}$ enhancement in the rare earth-doped
122-type iron-based pnictide superconductors {[}S. R. Saha et.al.
Phys. Rev. B \textbf{85}, 024525 (2012), arXiv:1105.4798 (2011); B.
Lv et al. PNAS \textbf{108}, 15705 (2011).{]}. Based on the real-space
effective c-axis lattice constant theory of superconductivity {[}X.
Q. Huang, arXiv:1001.5067{]}, it is shown that the abrupt c-axis reduction
of the superconductors is due to the structural phase transition (an
ultra-short-range symmetry breaking) of the charge stripe lattice.
The existence of this phase transition corresponds to the change from
the full-doped superconducting planes to the half-doped superconducting
planes in the studied superconductors. It is estimated that this phase
transition may help to promote the superconducting transition temperature
of the 122 family up to as high as 80 K.
\end{abstract}


\maketitle

\section{INTRODUCTION}

Twenty-six years after the discovery of high-temperature superconductivity
in copper-based superconductors \cite{bednorz,mkwu}, there is still
ongoing debate about how charge carriers move to maintain the superconducting
state in these materials \cite{anderson,chu}. Since the discovery
of iron-based superconductors \cite{kamihara}, researchers have expected
that the new superconductors may unlock the secrets of high-temperature
superconductivity. However, after five years' intensive study in the
iron-pnictide compounds \cite{zaren,xhchen,38K_122,43K,clarina},
it seems that the condensed matter physics community has become more
confused than ever. The researchers now find that they are unable
to answer the following fundamental question: Do the cuprate and iron-based
superconductors share an identical high-temperature mechanism of superconductivity?
Though it has been widely believed that there is a remarkable possibility
of an entirely different mechanism behind these two types of superconductors.
We firmly believe that an exactly the same intrinsic physical reason
is responsible for the superconductivity in both cases. Based on the
quantum confinement effect and the minimum energy principle, we have
proposed a unified description of cuprate and iron-based superconductivity
\cite{huang_c}.

\begin{figure}
\begin{centering}
\resizebox{1\columnwidth}{!}{ \includegraphics{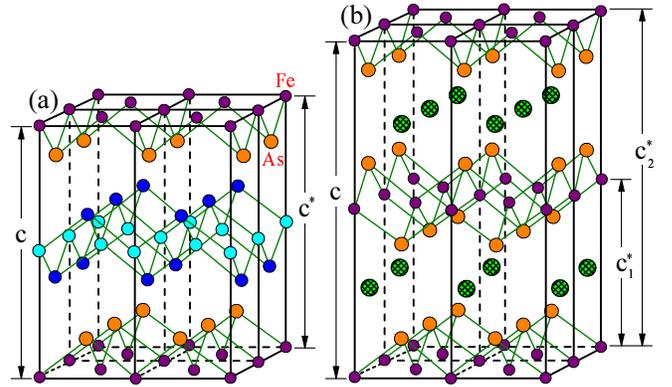}}
\par\end{centering}

\caption{Two types of the iron-based superconductors. (a) The 1111-type with
only single FeAs layer within one c-axis lattice constant, where the
c-axis lattice constant $c$ is equal to the effective c-axis lattice
constant $c^{*}$. (b) The 122-type having two FeAs layers within
the lattice constant $c$, in this case, there are two possible superconducting
phases: the full-doped low $T_{c}$ phase with the effective c-axis
lattice constant $c_{1}^{*}=c/2$ and the half-doped high $T_{c}$
phase of $c_{2}^{*}=c$. }

\label{fig1}
\end{figure}

The iron-based high-temperature superconductors started with the discovery
of superconductivity at 26 K in the 1111-type $LaFeAsO_{1\lyxmathsym{\textminus}x}F_{x}$
in 2008 \cite{kamihara}, very soon, the superconducting transition
temperature $T_{c}$ was raised to 55 K by replacing the lanthanum
with other rare-earth elements \cite{zaren}. Fig. \ref{fig1} shows
the two typical families of the iron-based superconductors, they are
1111-type \cite{kamihara} of Fig. \ref{fig1}(a) and 122-type \cite{38K_122}
of Fig. \ref{fig1}(b). Thus, facing the rapidly rising $T_{c}$,
some researchers claimed that room temperature superconductors made
possible from the iron-based material. At almost the same time, we
argued that the maximum $T_{c}$ of the 1111 family cannot exceed
60 K, while the 122 family has a maximum $T_{c}^{\max}<40$ K \cite{huang_maximum,huang_c}.
So far, our first prediction is still true by a number of experiments
tested in the past four years. But two different groups reported that
the $T_{c}$ of the rare earth-doped 122-type $CaFe_{2}As_{2}$ system
can be dramatically enhanced to more than 45 K \cite{saha,lv}, which
breakthrough the limitation of 40 K given by our theory in 2008. Does
this imply that our conclusion about the maximum transition temperature
of the 122-type iron-based superconductors is false?

Recently, the pressure-induced $T_{c}$ increase in $\beta-Fe_{1.01}Se$
compound was also observed \cite{medvedev}. Most recently, the collapsed
tetragonal phase of $Ca_{1-x}Pr_{x}Fe_{2}As_{2}$ was confirmed by
the NMR studies \cite{malong}. However, the nature of structural
collapse and the high-$T_{c}$ phase remains unclear. Here we will
show that, in the framework of effective c-axis lattice constant theory
of superconductivity \cite{huang_c}, the experimental observations
of structural collapse and $T_{c}$ enhancement in 122-type iron pnictide
superconductors can be well interpreted as a new ultra short-range
symmetry breaking of the real-space charge stripes. In this picture,
three different phase transitions (c-axis reduction, a-axis expansion
and $T_{c}$ enhancement) will occur simultaneously in the weak-doped
122-type superconductors. According to our theory, we conclude that
the maximum superconducting transition temperature of the collapsed
122 family may be promoted to about 80 K.

\section{WHAT IS THE KEY OF THE SUPERCONDUCTIVITY?}

By now, many theories and models have been developed to explain the
high-temperature superconductivity. As we all know, these theoretical
works have not been accepted by the scholarly consensus \cite{anderson}.
We fully agree with Anderson that many theories about electron pairing
in cuprate superconductors may be on the wrong track. In our opinion,
these theoretical studies did not take into account the most essential
point of the superconductivity. Then, what is the key of the high-temperature
superconducting phenomena?

\begin{figure}[t]
\begin{centering}
\resizebox{1\columnwidth}{!}{ \includegraphics{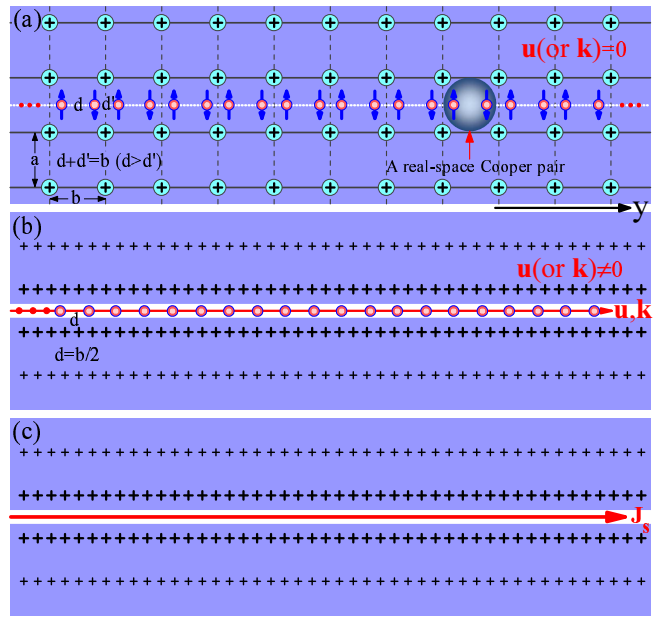}}
\par\end{centering}

\caption{(a) The real-space superconducting ground state where the electrons
self-assemble into some static one-dimensional peierls charge and
spin antiferromagnetic stripes, the real-space Cooper pair can be
formed inside single plaquette. (b) Under the influence of the external
fields, the peierls chains will transfer into periodic chains and
the electrons move at the same velocity $\mathbf{u}$ (or the same
momentum $\mathbf{k}$). (c) An equivalent model of (b), where the
superconducting current can flow without resistance along the ballistic
channel with a supercurrent density of $\mathbf{J_{s}}$.}

\label{fig2}
\end{figure}

From the interaction point of view, any superconducting system can
be simplified as a conventional classical electromagnetic interaction
system between the negative electrons and positive ion cores. From
the energy point of view, the superconducting state should be a stable
condensation state of electrons which do not radiate electromagnetic
energy. According to the classical electromagnetic theory, to maintain
a non-dissipative superconducting electronic state, the corresponding
superconducting electrons should not move with variable motion. In
other words, each superconducting electron can be considered as the
``inertial electron\textquotedblright{} on which the resultant force
is zero.

\begin{figure}
\begin{centering}
\resizebox{1\columnwidth}{!}{ \includegraphics{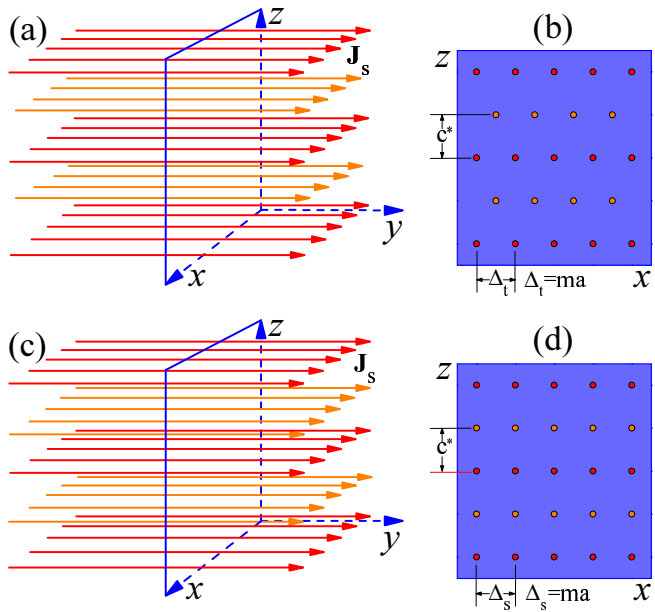}}
\par\end{centering}

\caption{The quasi-two-dimensional charge stripe lattices of different symmetries.
(a)-(b) triangle or equilateral triangle, (c)-(d) rectangular or square,
where $a$ is the a-axis lattice constant, $c^{*}$ is the effective
c-axis lattice constant defined in our theory and $m$ is a positive
integer.}

\label{fig3}
\end{figure}

In recent years, we have devoted considerable effort towards the development
of a unified theory of superconductivity \cite{huang_c,huang_maximum,magic}.
The new theory was established in the real-space picture, rather than
the commonly used momentum-space picture in superconducting areas
\cite{BCS}. In our approach, the most basic unit of the superconducting
ground state is the static one-dimensional charge chain (stripe) \cite{tranquada}
formed in the ab-plane of the crystal lattice. Without the external
field, it is not difficult to prove that the electrons will self-organize
into some quantized one-dimensional peierls chains with $d+d'=b$
($d>d'$), as shown in Fig. \ref{fig2}(a). Moreover, the real-space
Cooper pair is defined within one single plaquette, as indicated in
Fig. \ref{fig2}(a). Driven by the external fields, the ground state
of the peierls chain will spontaneously transform into one superconducting
excited state of a periodic chain with a definite electron-electron
spacing of $d=b/2$, as illustrated in Fig. \ref{fig2}(b). It is
obvious that all electrons are maintained in the zero-stress state
and move constantly at the same velocity \textbf{u} (or in the same
momentum \textbf{k}, which is differed from the BCS picture of opposite
momentum required for the formation of the Cooper pairs \cite{BCS}.).
The excited state of Fig. \ref{fig2}(b) can be equivalently described
in term of superconducting persistent-current $\mathbf{J}_{s}$, as
indicated in Fig. \ref{fig2}(c).

For the three-dimensional layered superconductors (where the c-axis
lattice constant $c>a$ and $c>b$), the charge stripes will undergo
further self-assembly into some quasi-two-dimensional stripe lattices,
as illustrated in Fig. \ref{fig3}. In this scenario, the superconducting
current can flow without resistance along some periodic array of ballistic
channels in the superconductor, as shown in Fig. \ref{fig3}(a) and
(c). Due to the symmetry of the stripe lattices {[}see Fig. \ref{fig3}(b)
and (d){]}, the stripe-stripe electromagnetic interactions can be
naturally suppressed to ensure the stable superconducting state. In
Fig. \ref{fig3}(b) and (d), the parameter $c^{*}$ is the so-called
effective c-axis lattice constant in our theory. Normally, the parameter
$c^{*}$ is taken to be the lattice constant $c$ of the superconducting
crystal. However, it may also be equal to $c/2$ in some special cases.
In the following, we will show that the parameter $c^{*}$ plays a
main role in determining the maximum $T_{c}$ of a given superconductor.

\section{THE MAXIMUM $T_{c}$ AND THE EFFECTIVE c-AXIS LATTICE CONSTANT}

It is our view that a successful theory of superconductivity can not
only explain the observed experiments, but it can also predict new
phenomena and give a clear physical description of the predicted results.
For a given superconducting material, perhaps the most difficult to
answer is: What is the highest superconducting transition temperature
expected for the superconductor? To the best of our knowledge, the
real-space effective c-axis lattice constant theory of superconductivity
can be considered as the first one that successfully estimate the
maximum $T_{c}$ of the studied compound.

In the framework of the lattice model of charge stripe (see Fig. \ref{fig3}),
the superconducting transition temperature of a superconductor is
closely related to the stability of the stripe lattice inside the
materials. In our theoretical model, the lattice vibrations always
have the tendency to destroy the existing superconducting state of
the stripe lattice. This implies that the BCS electron-phonon coupling
is probably not the cause of the superconductivity and the pairing
mechanism \cite{Hirsch}. Furthermore, we consider that the stripe-stripe
electromagnetic interaction is one of the most important factors that
affects the stability of the stripe lattice, in turn, influence the
maximum $T_{c}$ of the corresponding superconductor. Qualitatively,
a too strong stripe-stripe interaction could lead to a lower $T_{c}$
in the superconductor. Obviously, in order to promote the $T_{c}$
of the superconductor, it is necessary to control the Coulombic stripe-stripe
interaction to an optimum value.

With the help of Fig. \ref{fig3}, the most bewildering problem of
superconductivity in layered superconductors is no longer mysterious.
Now, the superconducting state is related simply to the ordered charge
structure due to the competition among the electrons. Because of the
intrinsic Coulomb repulsion between superconducting stripes, a superconducting
state in fact is an energy state with a condensed electromagnetic
energy which is called superconducting internal energy (SIE) in this
paper. In the following, we briefly discuss the relationship among
the SIE, the maximum $T_{c}$ and the lattice parameters of the studied
superconductor. Without losing the generality, we focus our discussion
on the striped triangular lattice of Fig. \ref{fig3}(b). In the first
approximation, the SIE can be expressed directly as

\begin{equation}
E_{SIE}=\frac{A(n,T)}{c^{*}}+\frac{B(n,T)}{ma},\label{SIE}
\end{equation}
where $n$ is the concentration of charge carriers (electrons), $T$
is the temperature, $A(n,T)$ and $B(n,T)$ are $n$ and $T$ related
constants, the parameters $c^{*}$, $a$ and $m$ are given in Fig.
\ref{fig3}.

It must be pointed out that Eq. (\ref{SIE}) is not suitable for the
following four extreme systems: (a) $n$ is too small, (b) $c^{*}$
is too large, (c) $n$ is too large and (d) $c^{*}$ is too small.
For the cases (a) and (b), the competitive interaction between electrons
is too weak to form the ordered superconducting stripe lattice. While
for the cases (c) and (d), the stripes are crowded and the stripe-stripe
interactions are too strong to allow a stable stripe lattice.

In our theoretical framework, for the doped high-temperature superconducting
materials, researchers can adjust the crystal structure of stripe
lattices by changing the electron concentration $n$, and thus alter
the internal energy $E_{SIE}$ and the superconducting transition
temperature $T_{c}$. Based on the relationship between symmetry and
stability of the system, the optimal doped sample corresponds to the
minimum energy of the stripe structure with regular triangle ($c^{*}=0.5\sqrt{3}\Delta_{t}$)
or square ($c^{*}=\Delta_{s}$) symmetry of Fig. \ref{fig3}. Hence,
for a high-temperature superconductor with an effective c-axis lattice
constant $c^{*},$ there exists a simple relation between $c^{*}$
and the minimum internal energy $E_{SIE}^{min}$:

\begin{equation}
E_{SIE}^{min}\propto\frac{1}{c^{*}}.\label{min}
\end{equation}

Here, the stripe lattice with the minimum energy $E_{SIE}^{min}$
corresponds to the maximum $T_{c}^{max}$ of a superconducting state.
The $T_{c}^{max}$ and $E_{SIE}^{min}$ satisfy the following relation:

\begin{equation}
T_{c}^{max}=A+\frac{B}{E_{SIE}^{min}}=A+Bc^{*},\label{Tmax}
\end{equation}
where $A$ and $B$ are two constants which can be determined experimentally.

\begin{figure}
\begin{centering}
\resizebox{1\columnwidth}{!}{ \includegraphics{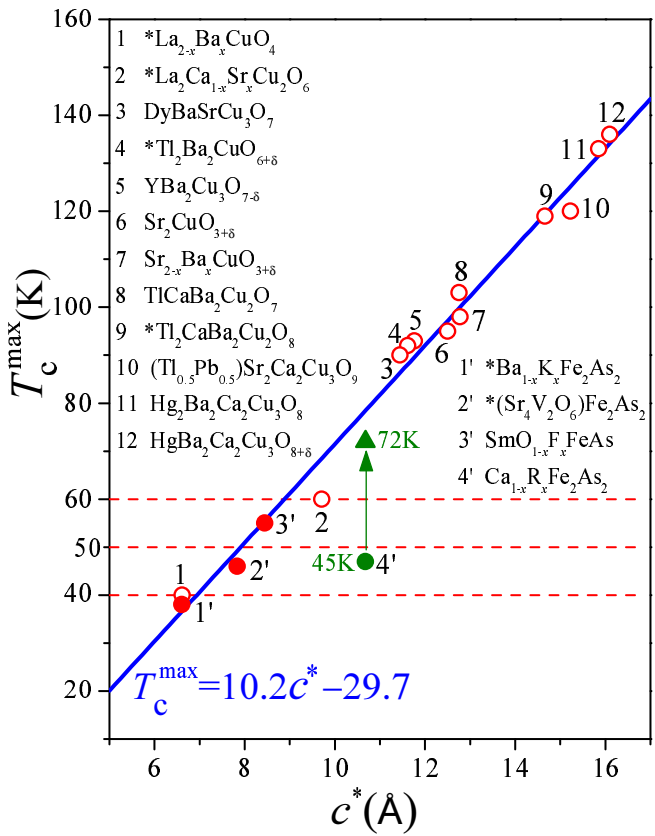}}
\par\end{centering}

\caption{The linear relationship between the measurements maximum $T_{c}$
and the effective c-axis lattice constant $c{}^{*}$ of Cu- and Fe-based
high-temperature superconductors. The blue line is obtained by fitting
the experimental data of copper superconductors (the hollow circles).
Here, the materials marked by ``{*}'' are those of $c{}^{*}=c/2$,
while the other satisfy $c{}^{*}=c.$ The green solid circle of 45
K is the experimental value of collapsed $Ca_{0.92}Nd_{0.08}Fe_{2}As_{2}$
sample and the triangle of 72 K is the expected value also from the
experiment of Saha et al. \cite{saha}.}

\label{fig4}
\end{figure}

It can be concluded from Eq. (\ref{Tmax}) that increasing the effective
c-axis lattice constant $c^{*}$ is the most effective way to raise
the maximum superconducting transition temperature of the high-temperature
layered superconductors. This conclusion is well confirmed by a large
number of experimental results reported for the cuprate and iron-based
superconductors, as shown in Fig. \ref{fig4}. In this figure, the
qualitative result of linear relationship of Eq. (\ref{TCmax}) between
$T_{c}^{max}$ and $c^{*}$ is clearly shown. It should be noted that
the fitted line is obtained with the fitting parameters of cuprate
superconductors (marked by twelve hollow circles) and the corresponding
equation is written as:

\begin{equation}
T_{c}^{max}\approx-29.7+10.2\times c^{*}.\label{TCmax}
\end{equation}

By comparing Eq. (\ref{TCmax}) with Eq. (\ref{Tmax}), the intercept
and slope of Eq. (\ref{Tmax}) are $A\approx-29.7$ and $B\approx10.2$,
respectively. Furthermore, we note that the experimental data of the
1111-, 21311- and 122-type iron-based superconductors also fall on
this line, as indicated by three solid circles in Fig. \ref{fig4}.
Based on this surprising result, we have hypothesized that the maximum
$T_{c}$ of the 1111-, 21311- and 122-type iron-based superconductors
cannot exceed 60 K, 50 K and 40 K (indicated by the dotted red lines
in Fig. \ref{fig4}), respectively. However, as pointed out in the
abstract, the limitation of 40 K has recently broken through in the
rare earth-doped 122-type $CaFe_{2}As_{2}$ systems \cite{saha,lv}.
In the next section, we will give an interpretation of these new experimental
results.

\section{STRUCTURAL COLLAPSE AND $T_{c}$ ENHANCEMENT IN 122-TYPE IRON-BASED
SUPERCONDUCTORS}

Why the structural collapse in 122-type family of iron pnictides can
result in a sharp increase of the $T_{c}$ value in the corresponding
samples? This question not only challenges the traditional understanding
of superconductivity, but also challenges our conclusion that the
maximum $T_{c}$ of the 122-type parent compounds cannot exceed 40
K. In this section, we show that the new experimental facts mentioned
above are the desirable results in our framework and they can be well
explained within the effective c-axis lattice constant theory of superconductivity.

According to our theory, an abrupt change of the superconducting transition
temperature is always associated with a radical change in the effective
c-axis lattice constant of the superconductor. Hence, in order to
explain the new phenomenon of $T_{c}$ enhancement one should start
from the lattice structure of Fig. \ref{fig1}(b), especially the
change of the effective c-axis lattice constant. Different from the
FeAs-1111 phase of Fig. \ref{fig1}(a), there are two superconducting
FeAs layers within one lattice constant $c$ in the FeAs-122 phase.
So in the 122-type family, there exist two different superconducting
phases. The first one is the full-doped phase where all the FeAs layers
are doped and contribute to the superconductivity, in this case, the
corresponding effective c-axis lattice constant satisfies $c_{1}^{*}=c/2$,
while the second one is the half-doped phase where the FeAs layers
doped interval with $c_{2}^{*}=c$, as indicated in Fig. \ref{fig1}(b).
Since $c_{2}^{*}>c_{1}^{*},$ so the half-doped phase naturally has
a higher $T_{c}^{max}$ than that of the full-doped phase.

Now the key issue is how to get the half-doped 122-type superconducting
samples in laboratory. Intuitively, a heavily doped (high carrier
concentration) sample tends to be in the half-doped phase, while the
light doped (low carrier concentration) sample may exist in the half-doped
structural state. These model predictions are consistent with the
experimental results, which indicated that the $T_{c}$ enhancement
occurs only at doping concentration $<16\%$. Next, we will present
a detailed analysis of the reported new physical phenomena by applying
our approach to the 122-type compounds.

\begin{figure}
\begin{centering}
\resizebox{1\columnwidth}{!}{ \includegraphics{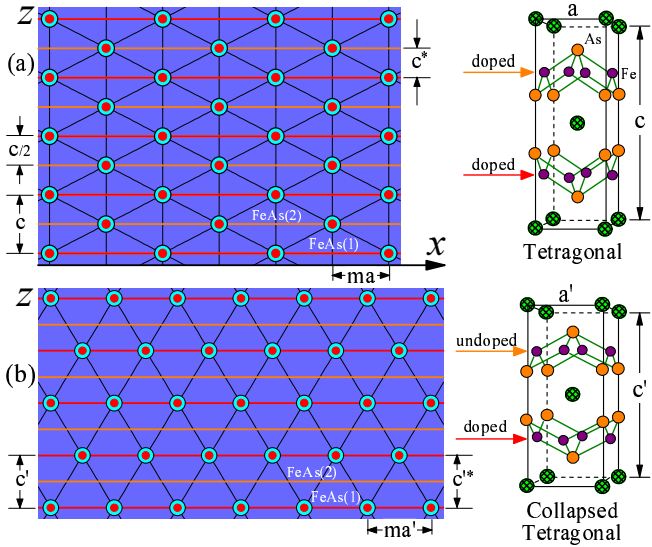}}
\par\end{centering}

\caption{The real space full-doped to half-doped charge-stripe phase transition
in the weak-doped 122-type FeAs superconductors. (a) The uncollapsed
full-doped phase, (b) the collapsed half-doped phase. This phase transition
will lead directly to the c-axis reduction ($c'<c$), a-axis expansion ($a'>a$) and at
the same time $T_{c}$ enhancement in the corresponding sample.}

\label{fig5}
\end{figure}

In 122-type FeAs superconductors, there are two sets of FeAs layers
which are FeAs(1) set and FeAs(2) set as illustrated in red and orange
lines respectively in Fig. \ref{fig5}. As can be seen from this figure,
the FeAs(1)-FeAs(1) or FeAs(2)-FeAs(2) spacing equals to the lattice
constant $c$, while FeAs(1)-FeAs(2) spacing is $c/2$. Generally,
both FeAs(1) and FeAs(2) layers are doped with charge carriers and
all FeAs layers are in the superconducting phase, as shown in Fig.
\ref{fig5}(a). In the absence of the external pressure, the charge
carriers (electrons) may enter into all FeAs layers and self-assemble
into a metastable full-doped phase. In this case, the corresponding
effective c-axis lattice constant is $c^{*}=c/2$ implying a relatively
small $T_{c}^{max}$ in this sample. However, when a pressure is applied
to the sample, the FeAs(1)-FeAs(2) spacing will decrease with the
increase of pressure. The shrinking of the spacing in turn could result
in a great increasing of the layer\textendash{}layer interaction,
while decreasing the instability of the lattice crystal and superconducting
stripe lattice.

When the applied pressure exceed the critical point, two structural
phase transitions of the crystal structure and stripe lattice structure
will happen simultaneously, as shown in Fig. \ref{fig5}(b). This
is the so-called full-doped to half-doped real space phase transition
which is driven by an external pressure. As we can see from Fig. \ref{fig5},
these phase transitions are closely related with the migration of
electrons from FeAs(2) layers to FeAs(1) layers, which can further
be viewed as an ultra short-range symmetry breaking of the stripe
lattice. This symmetry breaking will lead directly to the following
observable changes: (1) a dramatically abrupt c-axis contraction ($c\rightarrow c'$,
where $c/2<c'<c$) because of the decreasing of doped layer-layer
interaction ($c/2\rightarrow c'$) ; (2) a significant a-axis expansion
($a\rightarrow a'$, where $a<a'$) due to the increasing of in-plane
stripe-stripe interaction ($2ma\rightarrow ma'$) ; (3) an obvious
enhancement of $T_{c}$ originated from the increase of effective
c-axis lattice constant {[}$c^{*}(=c/2)\rightarrow c'^{*}(=c')${]}.
These qualitative analysis results are identical with the recent experimental
results \cite{saha,lv,ran}.

\begin{figure}
\begin{centering}
\resizebox{1\columnwidth}{!}{ \includegraphics{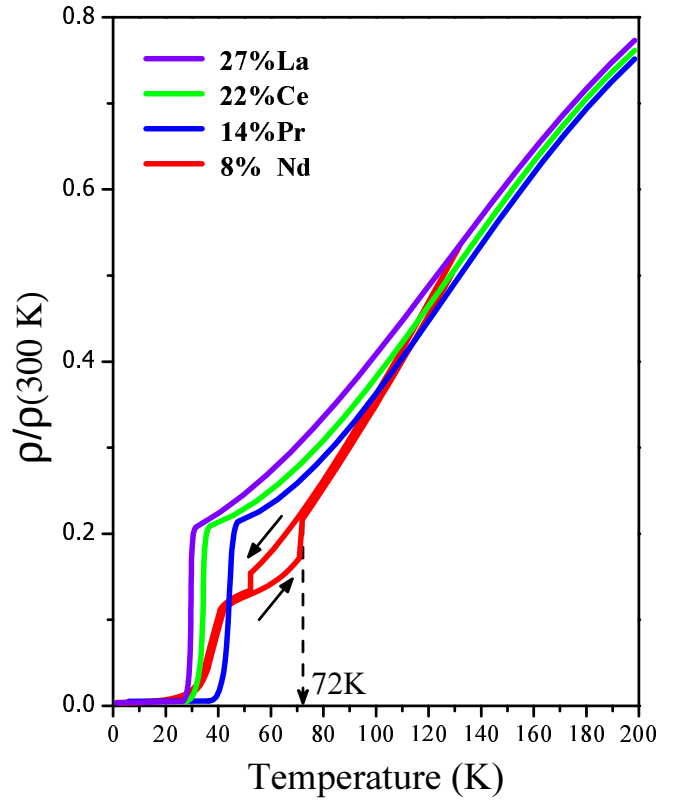}}
\par\end{centering}

\caption{The experimental results from Fig. 12 of Ref. \cite{saha} by Saha
et al. For the weak-doped sample of $Ca_{0.92}Nd_{0.08}Fe_{2}As_{2}$,
the corresponding superconducting transition temperature near 45 K
(the red line), the expected maximum $T_{c}^{max}$ is about 72 K
indicated by the dashed arrow. }

\label{fig6}
\end{figure}

According to the results of the previous section and the experimental
reported lattice constants \cite{saha}, it is possible to quantitatively
estimate the maximum $T_{c}^{max}$ in the collapsed 122-type iron-based
superconductors. In the paper \cite{saha}, the authors synthesized
a series of doped 122-type samples of different elements and different
concentrations, among which the sample $Ca_{1-x}Nd_{x}Fe_{2}As_{2}$
with $x$ around 0.09 was studied in detail. The crystallographic
data for this sample are ($a=3.9202\textrm{\AA}$, $c=11.273\textrm{\AA}$)
and ($a'=3.92822\textrm{\AA}$, $c'=10.684\textrm{\AA}$) for the
tetragonal structure and the collapsed tetragonal structure, respectively.
Moreover, the collapse transition temperature is around 80 K with
the uncertainty in temperature values being $\pm5$ K. Consequently,
the effective c-axis lattice constant of the collapsed sample is $c'^{*}=c'\approx10.684\textrm{\AA,}$
and then we can get the corresponding maximum $T_{c}^{max}\approx79.3$
K by putting the $c'^{*}$ value into Eq. (\ref{TCmax}). Note that
this theoretical value of $T_{c}^{max}$ is in complete coincidence
with the collapse transition temperature of 80 K, however, is much
higher than the experimental result of $T_{c}=45$ K for $Ca_{0.92}Nd_{0.08}Fe_{2}As_{2}$
, as shown in the red line in Fig. \ref{fig6}. We believe that the
difference between theory and experiment might be due to the quality
of the samples as well as to the non-optimal doping concentration.
In fact, from Fig. \ref{fig6} there is a clear evidence that superconducting
phase transition is most likely to occur at about 72 K which is very
close to the idea value of 79.3 K (see also Fig. \ref{fig4}), as
indicated in the dashed arrow in the figure.

\section{A BRIEF SUMMARY AND CONCLUSIONS}

Based on the real-space effective c-axis lattice constant theory of
superconductivity, we have studied the recent experimental findings
of structural collapse and $T_{c}$ enhancement in the rare earth-doped
122-type iron-based pnictide superconductors. We have argued that
the abrupt c-axis reduction of the superconductors is due to the full-doped
to half-doped structural phase transition (an ultra-short-range symmetry
breaking) of the charge stripe lattice. It has been shown that phase
transition directly leads to the expansion of effective c-axis lattice
constant, which in turn enhance the maximum superconducting transition
temperature of the studied materials. According to the experimental
crystallographic data, it has been estimated that the $T_{c}$ of
the collapsed 122 family can be raised to as much as 80 K by improving
the quality of single crystal component and by adjusting the carrier
concentration inside the superconductor. We have pointed out that
the three phase transitions (c-axis reduction, a-axis expansion, and
$T_{c}$ enhancement) observed in weak doped 122 series are intrinsically
correlated. In this paper, the physical nature of the new experimental
findings has been understood quite well on the basis of our approaches,
which implies that our framework may provide further insight into
the mechanism of the high-temperature superconductivity in general.

\end{document}